\documentstyle[prl,preprint,draft,aps,epsfig]{revtex}
\begin{document}
\title{Transverse optical Josephson plasmons, equations of motion}
\author{D. van der Marel, and A. A. Tsvetkov }
\address{Materials Science Center, Laboratory of Solid State Physics,
University of Groningen, }
\date{November 16, 2000}
\maketitle
\begin{abstract}
A detailed calculation is presented of 
the dielectric function in superconducttors consisting of 
two Josephson coupled superconducting layers per unit cell, taking 
into account the effect of finite compressibility of the electron 
fluid. From the model it follows, that two longitudinal, and one 
transverse optical Josephson plasma resonance exist in these materials,
for electric field polarization perpendicular to the planes. 
The latter mode appears as a resonance
in the transverse dielectric function, and it couples directly
to the electrical field vector of infrared radiation. A shift
of all plasma frequencies, and a reduction of the intensity of the
transverse optical Josephson plasmon is shown to result from the finite 
compressibility of the electron fluid. 
\end{abstract}
%
\section{Introduction}
In recent years the 'second' Josephson plasma resonance (JPR) phenomenon has 
been studied theoretically and has been observed 
experimentally\cite{pwa,lt21,shibata,munzar,ybco2jpr,bsco2jpr} in double layer high Tc cuprates. 
Originally the prediction of a transverse optical Josephson resonance and the
associated second JPR in layered superconductors with two layers per unit cell 
appeared as a short conference paper\cite{lt21}, explaining briefly the main
theoretical ingredients and results. One of the reasons for revisiting this 
problem is, that experimentally the transverse optical JPR was observed\cite{diana,dressel} 
in SmLa$_{0.8}$Sr$_{0.2}$CuO$_{4-\delta}$ with an
oscillator strength much smaller than expected on the basis of
the simple expression derived in Ref. \cite{lt21}. Following a
suggestion by Bulaevskii\cite{lev}, the expressions for the
dielectric function will be re-derived in this manuscript, while
taking into account an extra term representing the finite compressibility 
of the electron fluid. The central result is, that the dielectric function is 
\begin{displaymath}
  \frac{1}{\epsilon(\omega)} =  
    \frac{\tilde{z_I}\omega^2}{\omega^2-\tilde{\omega}_I^2} + 
    \frac{\tilde{z_K}\omega^2}{\omega^2-\tilde{\omega}_K^2}                                   
  \end{displaymath}
which is the same expression as in Ref. \cite{lt21}, except that the volume fractions 
$x_I$ and $x_K$ have been replaced with {\em effective} weight factors 
$\tilde{z}_I$ and $\tilde{z}_K=1-\tilde{z}_I$, which depend on the volume fractions,
the plasma frequencies $\omega_I$ and $\omega_K$, and the compressibility. 
\\
Let us consider the optical response function for a
material with two superconducting layers per unit cell. The lattice constant
along the c-direction, perpendicular to the planes, is $d$.  
The layers are grouped in pairs, with interlayer distances $x_I d$ and $x_K d$
alternating ($x_I+x_K=1$). Hence the z-coordinate of the m'th plane normalized by the
lattice constant is $x_m = m/2$ if $m$ is even, and $x_m=(m-1)/2+x_I$ if
$m$ is odd.                                  
\\
The charge fluctuation of each plane is characterized by a charge amplitude $Q_m$ 
and a phase $\phi_m$, where $m$ is the layer index, and $d$ is
the length of the unit cell along the c-direction. The discussion in this paper
will be restricted to
the charge fluctuations perpendicular to the planes, corresponding
to a homogeneous charge distribution within each plane. In this case the electric
fields are perpendicular to the planes. The electric field of a single plane with an 
area $A$, and charged with positive charge $Q$ is constant in space, and the field
lines are directed away from the plane. As a result the potential energy of a positive
test charge with charge $q$ in this field {\em de}creases linearly as a function of the
distance $d$ from the plane: $E_{pot}(z)=-4\pi qQ z /A$. The potential energy 
stored in the field of the charge fluctuations $Q_m$ and $Q_n$ of a pair of planes 
at a distance $d|x_{m}-x_{n}|$ is half this amount 
\begin{equation}
 V_C(Q_m,Q_n) = -\frac{|x_{m}-x_{n}|Q_mQ_n}{2C_0}
 \label{ecoul}
\end{equation}
where $C_0=A/(4\pi d)$ is the capacitance of two planes at a distance corresponding to the
lattice parameter $d$. The total potential energy stored in the fields of
the charge fluctuations is just the linear superposition of the contributions from
all pairs of planes in the crystal. In this
context it is important to point out, that the planes considered here
are truly two-dimensional (2D) in the electrodynamical sense: Because the charge
has no spatial degrees of freedom perpendicular to the planes (other than
tunneling between planes), the individual planes provide no channel for
metallic screening for fields
along the c-axis. This is quite different from the situation encountered in classical
Josephson junctions between thick metallic layers. In the latter case individual
planes {\em do} screen the electric fields polarized along the c-direction. 
\\ 
The second source of potential energy is the
the electronic compressibility. This is due to
the fact that if $\delta N$ electrons are 
added to a plane, the free energy increases with an amount 
$\delta F = \mu \delta N + \delta N^2/(2K_0n^2)$, where 
$\mu$, $K_0$, and $n$ are the chemical potential, the electronic
compressibility, and the electron density respectively. For a
Fermi-gas $K_0n^2=\partial n/ \partial \mu$ corresponds to 
density of states at the Fermi level. 
In the context of 'excitons' in two-band superconductors, the 
compressibility term has been first considered in 1966 by 
Leggett\cite{tony}. In neutral fluids the compressibility causes 
propagation of sound, whereas for electrons it causes the dispersion 
of plasmons. Plasma dispersion of the JPR in the cuprates has been 
described by Koyama and Tachiki\cite{koyama}. In part the compressibility 
can be motivated by calculations based on the random
phase approximation(RPA), showing a finite dispersion of 
the charge density fluctuations in single layer\cite{fertig} and
bilayer\cite{hwang} cuprates, and of spin, amplitude and phase 
collective modes\cite{vdmarel}. It has been shown by Artemenko and Kobel'kov, 
that the frequency of the resonance, its
dispersion and its damping are strongly influenced by the presence of
quasi-particles at finite temperature\cite{artemenko}. Although 
this type of calculation demonstrates, that the Pauli exclusion principle causes a
a finite compressibility of the electron fluid, leading to finite dispersion of the
collective modes, weak coupling approaches are not well supported due to the strong 
electronic correlations in these materials. We thefore treat the electronic 
compressibility as a phenomenological parameter in this paper.
\\
The terms in the free 
energy proportional to $\delta N$ only shift the equilibrium density. 
In  harmonic approximation, the charge fluctuations around equilibrium 
follow from the quadratic terms
\begin{equation}
  V_K(Q_m) = \frac{(Q_m/e)^2}{2 K n^2 A} =  \frac{\gamma_0}{2C_0} Q_m^2
  \label{compress}
\end{equation}
For later convenience the dimensionless constant 
$\gamma_0 = 1/ (4\pi d e^2 K_0n^2)$, proportional to the 2D bulkmodulus, 
is introduced here, to characterize the compressibility.
In Ref. \cite{lt21} we left these terms 
out of consideration. 
\\
We calculate the longitudinal dielectric function following the usual procedure of
adding external charges $Q^{e}_m$ to each layer, distributed such as to provide an
external electric field $D$ (the displacement field) of the plane wave form
with wavevector $k$. The {\em definition} of the dielectric constant\cite{mahan} 
implies, that the internal and external charge distributions interact only 
via the electromagnetic field. Hence, the interaction between internal and external 
charge is described by the Coulomb term, Eq. \ref{ecoul}, but the external charge 
does not enter the compressibility term, Eq. \ref{compress}.  
The charge dynamics enters via the Josephson coupling $J_{m}^{m+1}$ between 
each set of nearest neighbor planes
\begin{equation}
  H_{kin} = -J_{m}^{m+1}\cos(\phi_m-\phi_{m+1})
\end{equation}
Our aim is to determine the
dielectric constant, and collective modes in the absence of external 
DC magnetic fields. For this purpose we will need the equations 
of motion for the {\em internal} charge accelaration $d^2Q^{i}_m/dt^2$ 
subject to the fields of the internal {\em and} external charges 
$Q^{t}_m=Q^{i}_m + Q^{e}_m$. These  follow from the Hamiltonian
\begin{equation}
  H = -\sum_{m>n}\frac{|x_m-x_n|}{2C_0} (Q^i_m+Q^e_m)(Q^i_n+Q^e_n)
  +  \sum_{m}\frac{\gamma_0}{2C_0} (Q^i_m)^2 
  -\sum_m J_{m}^{m+1}\cos(\phi_m-\phi_{m+1})
 \label{hamiltonian}
\end{equation}
Here the phases $\phi_m$ and the internal charges $Q^i_m$ are conjugate variables, 
which are subject to the Hamilton-Josephson equations of motion:
$\frac{\hbar}{e^*}d\phi_m/dt=\partial H/\partial Q^i_m$, and
$\frac{\hbar}{e^*}dQ^i_m/dt=-\partial H/\partial \phi_m$,
where $e^*=2e$ is the charge of a Cooper-pair. 
\section{Equations of motion}
Working in the linear response
regime, we obtain the equations of motion for the {\em internal} charge
accelaration $d^2Q^{i}_m/dt^2$ subject to the fields of the internal {\em and}
external charges 
\begin{equation}
 \begin{array}{l}
   - 2C_0\left\{\frac{\hbar}{e^*}\right\}^2 \frac{d^2Q^{i}_m}{dt^2} =  
  \\ 
   -\sum_{n}\left\{\left[J_{m}^{m+1}+J_{m-1}^{m}\right]|x_n-x_m|
    -J_{m}^{m+1}|x_n-x_{m+1}| -J_{m-1}^{m}|x_n-x_{m-1}\right|\}Q^{t}_n    
  \\ 
    + 2\gamma_0 (J_{m}^{m+1}+J_{m-1}^{m})Q^i_m - 2\gamma_0 J_{m}^{m+1}Q^i_{m+1} 
   - 2\gamma_0 J_{m-1}^{m}Q^i_{m-1}
 \end{array}                                               
 \label{eom}
\end{equation}
Due to the lattice periodicity, the solutions have to be plane waves with a 
wavevector $k=\phi/d$. Therefor we can use generalized charge coordinates 
$R$, and $S$ defined as $Q_{2m} = S e^{im\phi}$, $Q_{2m+1} = R e^{im\phi}$, 
to describe the charge fluctuations in the even and odd planes. The Josephson 
coupling energies $J_{2m}^{2m+1}=I$ and $J_{2m-1}^{2m}=K$ characterize the two 
types of junctions. 
\\
It is quite easy to extend this to the situation, where we have a lattice polarizability
characterized by a dielectric constant $\epsilon^s_I$ and 
$\epsilon^s_K$ for each type of Josephson junction. 
This corresponds to the transformation  
$C_0 \rightarrow C_{av}$, $\gamma_0 \rightarrow \gamma$,
$x_I \rightarrow z_I$, and $x_K \rightarrow z_K$
in Eqs. \ref{hamiltonian}, and \ref{eom}. Here the following definitions
have been used:
$C_{av}= A\epsilon^s_{av}/4\pi d$ for the average capacitance,
$\gamma=\epsilon_{av}/ (4\pi d e^2 K_0n^2)$ for the compressibility,
$z_I=x_I\epsilon^s_{av}/\epsilon^s_I$ and $z_K=x_K\epsilon^s_{av}/\epsilon^s_K$
(together satisfying $z_K+z_I=1$) for the weight factors, 
and $1/\epsilon^s_{av}=x_K/\epsilon^s_K+x_I/\epsilon^s_I$ for the average dielectric constant.
We are now ready to formulate the equations of motion for the generalized coordinates 
$S$ and $R$ for each wavenumber $k=\phi/d$.  
\begin{equation}
 \begin{array}{l}
    2C_{av}\left\{\frac{\hbar\omega}{e^*}\right\}^2 S^{i} = 
   -\sum_{n} e^{i\phi n}
   \left\{ |n| (I+K) - |n-z_I| I -|n+z_K| K \right\} S^{t}  \\  
   - \sum_{n} e^{i\phi n}\left\{ |n+z_I| (I+K) - |n| I -|n+1| K \right\} R^{t}     
   + 2\gamma(I + K)S^i -  2\gamma(I + e^{-i\phi}K)R^i       \\
   \\
    2C_{av}\left\{\frac{\hbar\omega}{e^*}\right\}^2 R^{i} = 
   - \sum_{n} e^{i\phi n}\left\{ |n| (I+K) - |n+z_I| I -|n-z_K| K \right\} R^{t}  \\  
   - \sum_{n} e^{i\phi n}\left\{ |n-z_I| (I+K) - |n| I -|n-1| K \right\} S^{t}    
   +2\gamma(I + K)R^i -  2\gamma(I + e^{i\phi}K)S^i      \\      
 \end{array}                                               
 \label{eom2}
\end{equation}
The convergent lattice sums over $n$ can be replaced with the identities
\begin{equation}
 \sum_n|n+p \pm x|e^{in\phi}=e^{-ip\phi}\frac{(1- e^{\mp i\phi})x-1}{1-\cos\phi}                                               
 \label{defs2}
\end{equation}
Before we continue it is convenient to partly diagonalize 
Eq. \ref{eom2}, by transforming the expressions to new generalized 
charge coordinates $Q$ and $P$, defined as $Q=S+R$ and $P=e^{i\phi}S+R$, with
identical transformations for the internal and external charges.
We will see later, that $Q$ and $P$ correspond to charge fluctuations across 
the barriers of type $K$ and $I$ respectively. The reverse transformations are
$S=(Q-P)/(1-e^{i\phi})$, and $R=(P-e^{i\phi}Q)/(1-e^{i\phi})$.  
In addition it will turn out to be convenient to introduce
the Josephson plasma resonance frequencies
characteristic of the two types of junctions 
$\omega_{K}^2 = z_K K (e^*/\hbar)^{2} C_{av}^{-1}$, and 
$\omega_{I}^2 = z_I I (e^*/\hbar)^{2} C_{av}^{-1}$, and the corresponding local
charge response functions $\epsilon_K=1-{\omega_{K}^2}/{\omega^2}$, 
and $\epsilon_I=1-{\omega_{I}^2}/{\omega^2}$. With the help of the identities 
Eq. \ref{defs2} the equations of motion of the external and total generalized charge
coordinates become
\begin{equation}
 \begin{array}{lll}
   Q^{e} &=& 
   \epsilon_K Q^t - 2\gamma z_K^{-1}\omega_{K}^2\omega^{-2}(Q^i-P^i e^{-i\phi/2}\cos{(\phi/2)}) \\  
   P^{e} &=& 
   \epsilon_I P^t - 2\gamma z_I^{-1}\omega_{I}^2\omega^{-2}(P^i-Q^i e^{i\phi/2}\cos{(\phi/2)}) \\  
 \end{array}                                               
 \label{eom6}
\end{equation}
We see, that if the compressibility term $\gamma=0$, these equations of motion are already
diagonal, corresponding to two non-dispersing longitudinal plasmons at frequencies
$\omega_K$ and $\omega_I$. It is also immediately clear from this, that $Q$, and $P$ 
correspond to the charge fluctuations across junctions of type $K$, and $I$ respectively.
If $\gamma$ is finite, the equations of motion are coupled, and the plasma frequencies
will have a finite dispersion as a function of $k=\phi/d$.
\section{Calculation of the dielectric function}
We are interested in the response of the total electric field $E$ to an
external field $D$ which is polarized along the c-axis, and which
varies harmonically in time and space along the $c$-direction, {\em i.e.}
\begin{equation}
  \vec{D}(\vec{r},t) =  D_0\hat{z}e^{i(k z-\omega t)}                                   
\end{equation}
The dielectric function $\epsilon(k,\omega)$ is calculated from the definition
$\vec{D}=\epsilon\vec{E}$. We will employ the fact, that $\vec{D}=\vec{\nabla} V^e$
and $\vec{E}=\vec{\nabla} V$. $V$, $V^i$, and $V^e$ are the total, internal and the
external voltages respectively. Thus we need to arrange the external charges 
in a such a way, as to garantee that the external
field $D(z)$ has a plane wave form. For $z$ coinciding with the coordinates of 
the conducting planes, this requires that 
${V_{2m+1}^e}/{V_{2m}^e} = e^{iz_I\phi} $                    
We satisfy this requirement by giving the external charges the ratio
${R^e}/{S^e}=\{e^{iz_I\phi}-z_K-z_Ie^{i\phi}\}/\{1-z_Ke^{iz_I\phi}-z_Ie^{-iz_K\phi}\}$.
%
%
For the generalized coordinates this implies
\begin{equation}
  \frac{P^e}{Q^e}=\frac{z_K\sin(z_I\phi/2)}{z_I\sin(z_K\phi/2)}e^{i\phi/2}                
 \label{testB}
\end{equation}
For the calculation of $\epsilon$ we need to calculate the macroscopic average of $E$, and $D$,
corresponding to the macroscopic electric and displacement fields. For this it is sufficient to
know the voltages at the positions of the planes. Now, using Eq. \ref{ecoul} we observe, that 
the voltages in each plane are
\begin{equation}
  \begin{array}{lll}
  V_{2m} &=& \frac{-1}{C_{av}}\sum_n\left(Q_{2n}|n-m|+Q_{2n+1}|n-m+z_I|\right)  \\
  V_{2m+1} &=& \frac{-1}{C_{av}}\sum_n\left(Q_{2n}|n-m-z_I|+Q_{2n+1}|n-m|\right)  \\
  \end{array}                                  
 \label{VA}
\end{equation}
We consider a charge oscillation with wavevector $k=\phi/d$. The charges in the
alternating layers are $Q^{t}_{2n}=e^{in\phi}S^{t}$ and
$Q^{t}_{2n+1}=e^{in\phi}R^{t}$. After summation over the index $n$, the
voltages in the zeroth and first plane 
\begin{equation}
  \begin{array}{llllll}
  V_{0} &=& \frac{-1}{C_{av}(\cos\phi-1)}\{S^{t}+R^{t}(z_K+z_Ie^{-i\phi})\} \\
  V_{1} &=& \frac{-1}{C_{av}(\cos\phi-1)}\{S^{t}(z_K+z_Ie^{i\phi})+R^{t}\} \\
  \end{array}                                  
 \label{VB}
\end{equation}
with similar expressions for the external charges,for which $C_{av}$ must be replaced 
with $C_{0}$. The electric fields integrated between the nearest neighbor planes are
\begin{equation}
 \begin{array}{l}
 \int_{(m-z_K)d}^{md} E(z) dz =  V_{2m} - V_{2m-1} = C_{av}^{-1}e^{im\phi}(e^{-i\phi}-1)z_K Q^{t} \\
 \int_{md}^{(m+z_I)d} E(z) dz =  V_{2m+1} - V_{2m} = C_{av}^{-1}e^{im\phi}(e^{-i\phi}-1)z_I P^{t} \\
 \end{array}                                  
 \label{VA}
\end{equation}
with similar expressions for D, for which $C_{av}$ must be replaced with $C_{0}$. In 
the limit $k\rightarrow 0$ the macroscopic
electric field is just the sum of the two integrals devided by the lattice 
parameter $d$. We conclude from this, that for $k\rightarrow 0$ the dielectric 
function is
\begin{equation}
  \epsilon(\omega) = \frac{\int_{0}^{d} D(z) dz}{\int_{0}^{d} E(z) dz}
  =\epsilon_{av}\frac{z_K Q^e+z_I P^e}{z_K Q^{t}+z_I P^{t}}                                    
 \label{epsilon2}
\end{equation}
From Eq.\ref{testB} we see, that $P^e=Q^e$ in the limit $k\rightarrow 0$.
We can combine this identity with the equations of motion, Eq. \ref{eom6}, to prove that
${P^{t}}/{Q^{t}} = \{\epsilon_K - 2\gamma({\omega_K^2}/{z_K}+{\omega_I^2}/{z_I})\omega^{-2}\}
 /\{\epsilon_I - 2\gamma({\omega_K^2}/{z_K}+{\omega_I^2}/{z_I})\omega^{-2}\}$,
that
${P^e}/{P^{t}} = \epsilon_I-
 \{(\epsilon_K-\epsilon_I)2\gamma\omega_I^2/z_I\}/
   \{\epsilon_K\omega^2-2\gamma(\omega_K^2/z_K+\omega_I^2/z_I)\}$,
and
${Q^e}/{Q^{t}} = \epsilon_K-
 \{(\epsilon_I-\epsilon_K)2\gamma\omega_K^2/z_K\}/
   \{\epsilon_I\omega^2-2\gamma(\omega_K^2/z_K+\omega_I^2/z_I)\}$.
The dielectric function is now easily obtained,
\begin{equation}
  \frac{1}{\epsilon(\omega)} = \frac{1}{\epsilon_{av}}
  \frac{\omega^2\left(\omega^2-\tilde{\omega}_T^2\right)}
  {(\omega^2-\tilde{\omega}_I^2)(\omega^2-\tilde{\omega}_K^2)}
 \label{epsilon5}
\end{equation}
with the definitions
\begin{equation}
 \begin{array}{lll}
  \tilde{\omega}_K^2&=&(\frac{1}{2}+\frac{\gamma}{z_K})\omega_{K}^2+
  (\frac{1}{2}+\frac{\gamma}{z_I})\omega_{I}^2+ 
  \sqrt{\left((\frac{1}{2}+\frac{\gamma}{z_K})\omega_{K}^2
     -(\frac{1}{2}+\frac{\gamma}{z_I})\omega_{I}^2\right)^2
   +\frac{(2\gamma\omega_{K}\omega_{I})^2}{z_Kz_I}}    
  \\
  \tilde{\omega}_I^2&=&(\frac{1}{2}+\frac{\gamma}{z_K})\omega_{K}^2+
  (\frac{1}{2}+\frac{\gamma}{z_I})\omega_{I}^2- 
  \sqrt{\left((\frac{1}{2}+\frac{\gamma}{z_K})\omega_{K}^2
     -(\frac{1}{2}+\frac{\gamma}{z_I})\omega_{I}^2\right)^2
   +\frac{(2\gamma\omega_{K}\omega_{I})^2}{z_Kz_I}}      
  \\ 
  \tilde{\omega}_T^2&=&
  (z_I+\frac{2\gamma}{z_K})\omega_K^2+(z_K+\frac{2\gamma}{z_I})\omega_I^2
  \\
  \end{array}                                  
 \label{tildexy}
\end{equation}
Here we have adopted the convention for labeling the two plasma-resonances such, 
that $K$ always refers to the highest plasma resonance frequency.
\section{Central result}
Provided that $\tilde{\omega}_I<\tilde{\omega}_T<\tilde{\omega}_K$, it is always possible 
to express $\tilde{\omega}_T$ as a weighted average of the two longitudinal frequencies
\begin{equation}
  \tilde{\omega}_T^2=\tilde{z}_K\tilde{\omega}_I^2+\tilde{z}_I\tilde{\omega}_K^2
\end{equation}
with weight factors satisfying $\tilde{z}_K+\tilde{z}_I=1$. The latter
are no longer the volume fractions $z_I$ and $z_K$, as in Ref.\cite{lt21}. Instead
they depend on the volume fractions, {\em and} on the microscopic electronic 
parameters characterizing the two types of junctions. The effective fractions
can be calculated by inverting the above relation, {\em i.e.} using 
\begin{equation}
 \tilde{z}_K= (\tilde{\omega}_K^2-\tilde{\omega}_T^2)/(\tilde{\omega}_K^2-\tilde{\omega}_I^2)
 \label{zks1}
\end{equation}
As a result\cite{howepsilon6} we can write the inverse dielectric function, Eq.\ref{epsilon5} as a 
linear superposition of two plasma resonances
\begin{equation}
  \frac{\epsilon_{av}}{\epsilon(\omega)} =  
  \frac{\tilde{z_I}\omega^2}{\omega^2-\tilde{\omega}_I^2} + \frac{\tilde{z_K}\omega^2}{\omega^2-\tilde{\omega}_K^2}                                   
 \label{epsilon6}
\end{equation}
which is the same expression as in Ref. \cite{lt21}, except that
$z_I$ and $z_K$ have been replaced with {\em effective} volume fractions 
\begin{equation}
  \tilde{z}_K= \frac{1}{2}   
    \pm \frac{(z_K-z_I)(z_Kz_I+2\gamma)}{2(z_Kz_I+2\gamma+4\gamma^2)}
    \sqrt{1-\frac{4(2\gamma)^2\tilde{\omega}_K^2\tilde{\omega}_I^2}
         {(z_Kz_I+2\gamma)(\tilde{\omega}_K^2-\tilde{\omega}_I^2)^2}} 
    - \frac{2\gamma(z_Kz_I+\gamma)}{z_Kz_I+2\gamma+4\gamma^2}
      \frac{\tilde{\omega}_K^2+\tilde{\omega}_I^2}{\tilde{\omega}_K^2-\tilde{\omega}_I^2}                                  
 \label{tildez}
\end{equation}
Together, Eqs. \ref{epsilon6} and \ref{tildez} form the central result of this paper. 
From the intensities of the experimental loss functions the effective volume fractions 
$\tilde{z}_K$ and $\tilde{z}_I$ can be extracted. These can be used to calculate $\gamma$, 
and this in turn can be used to determine the density of states
\begin{equation}
  \frac{\partial n}{\partial\mu} =  \frac{1}{4\pi d e^2}\frac{1}{\gamma}
  \label{kn2}
\end{equation}
or, using $K_0n^2=\partial n/ \partial \mu$, the compressibility.
Once $\tilde{\omega}_I$ and $\tilde{\omega}_K$ have been measured, and $\gamma$ has been 
calculated from the weight factors $\tilde{z_K}$ and $\tilde{z_I}$ using Eq. \ref{tildez},
it becomes possible to make two further deductions, namely the determination of 
$\omega_K$ and $\omega_I$ using
\begin{equation}
 \begin{array}{lll}  
  \omega_{K}^2 &=& 
  \frac{z_K}{z_K+2\gamma}
  \left\{ 
   \frac{ \tilde{\omega}_K^2+\tilde{\omega}_I^2 }{ 2 }  
    \pm
   \frac{ \tilde{\omega}_K^2-\tilde{\omega}_I^2 }{ 2 }
     \sqrt{
        1-\frac{ 4(2\gamma)^2\tilde{\omega}_K^2\tilde{\omega}_I^2 }
          { (z_Kz_I+2\gamma)(\tilde{\omega}_K^2-\tilde{\omega}_I^2)^2 }} \right\}
  \\
 \omega_{I}^2 &=&
 \frac{z_I}{z_I+2\gamma}
  \left\{ \frac{\tilde{\omega}_K^2+\tilde{\omega}_I^2}{2} 
  \mp
  \frac{\tilde{\omega}_K^2-\tilde{\omega}_I^2}{2}
  \sqrt{1-\frac{4(2\gamma)^2\tilde{\omega}_K^2\tilde{\omega}_I^2}
         {(z_Kz_I+2\gamma)(\tilde{\omega}_K^2-\tilde{\omega}_I^2)^2}} \right\}
  \\
  \end{array}                                  
 \label{IK}
\end{equation}
from which we can calculate directly the Josephson coupling
energies
\begin{equation} 
 \begin{array}{lll}                               
   K = \frac{\epsilon_{av}}{4\pi d} \left\{\frac{\hbar\omega_{K}}{e^*}\right\}^2 
   &\mbox{and}&
   I = \frac{\epsilon_{av}}{4\pi d} \left\{\frac{\hbar\omega_{I}}{e^*}\right\}^2\\
 \end{array}
 \label{Ejos}
\end{equation}
\section{Evolution of the oscillator strenght as a function of $\gamma$}
\begin{figure}[h!]
  \centerline{\epsfig{figure=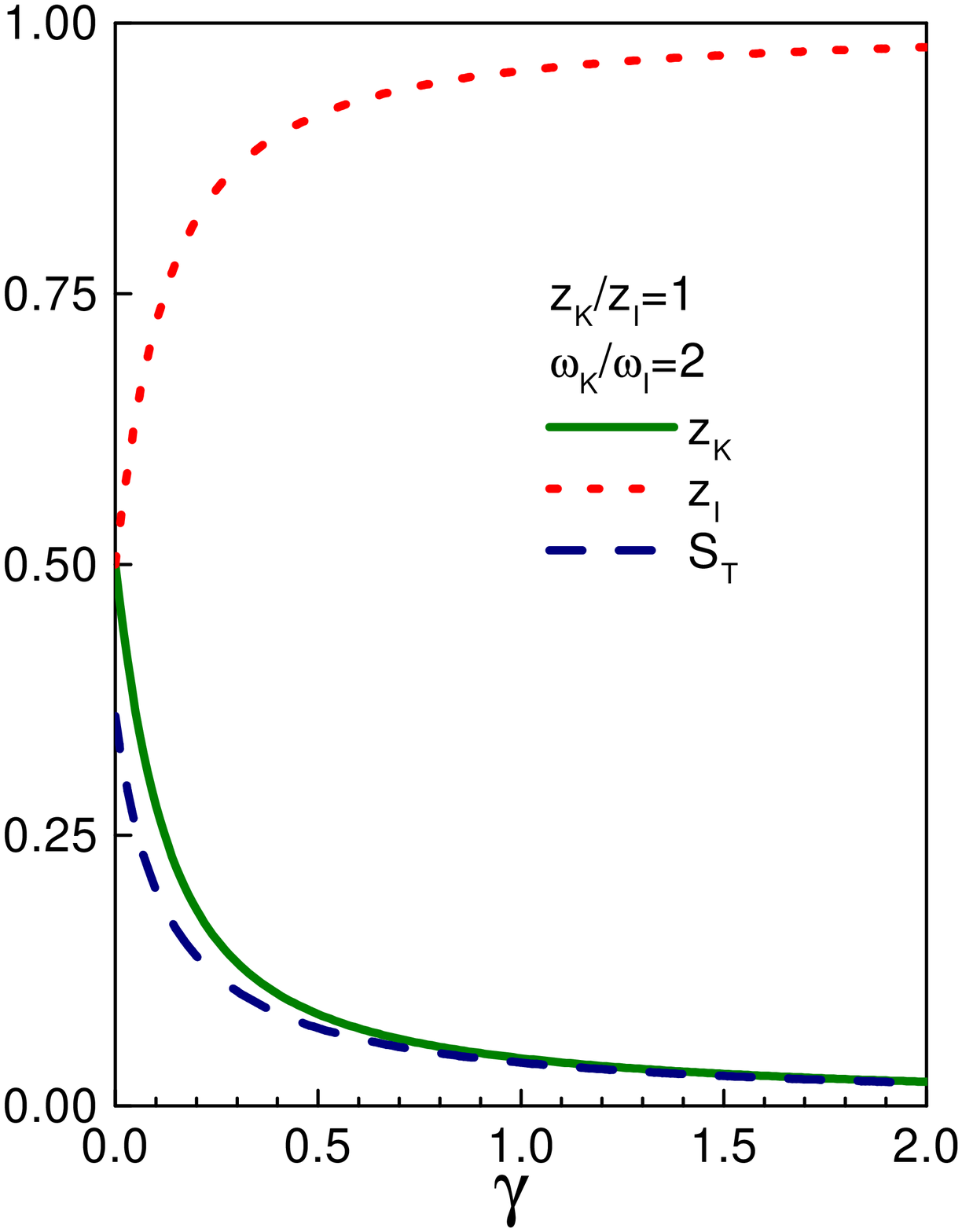,width=7cm,clip=}
  \epsfig{figure=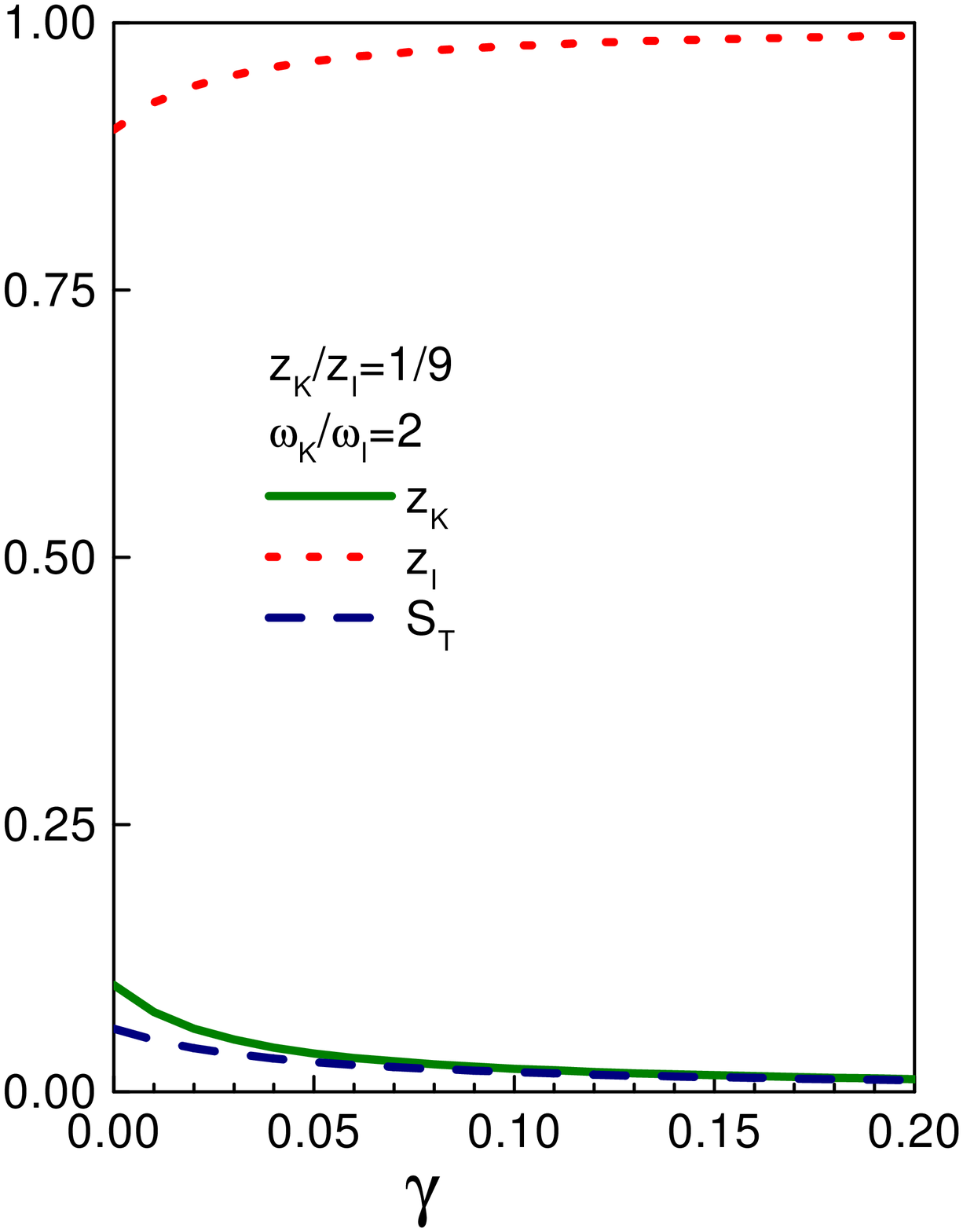,width=7cm,clip=} } 
 \caption{Oscillator strength's as a function of $\gamma$. }
 \label{fig:strength}
\end{figure}

For the analysis of experimental data Eqs. \ref{epsilon6} and \ref{tildez} suffice to
deduce the microscopic parameters $I$, $K$, and $K_0n^2$, {\em i.e.} the two
Josephson energies and the compressibility factor. To predict the plasma
resonance frequencies $\tilde{\omega}_T$, $\tilde{\omega}_I$, and $\tilde{\omega}_K$,
we can use Eq. \ref{tildexy}. The intensities of the two peaks in the energy loss 
function, Im$-1/\epsilon(\omega)$, are just the weight factors $\tilde{z_K}$ and $\tilde{z_I}$.
Their dependence on the microscopic parameters $\omega_I$, $\omega_K$, and $K_0n^2$ is given by the
expressions
\begin{equation}
  \tilde{z}_K=1-\tilde{z}_I= \frac{1}{2} +
  \frac{(z_K-z_I)(\omega_K^2-\omega_I^2)-2\gamma(\omega_K^2/z_K+\omega_I^2/z_I)}
  {2\sqrt{\left((1+\frac{2\gamma}{z_K})\omega_{K}^2
     -(1+\frac{2\gamma}{z_I})\omega_{I}^2\right)^2
   +4\frac{(2\gamma\omega_{K}\omega_{I})^2}{z_Kz_I}}}                                   
 \label{tildez2}
\end{equation}
The compressibility is characterized by the dimensionles parameter $\gamma$. The 
most important effect of this extra term is, that the intensity of the highest 
plasma resonance, $\tilde{\omega}_K$, is reduced compared to what it would have 
been if $\gamma$ were zero: It is clear from Eq. \ref{tildez2}, that the effective volume 
fraction $\tilde{z_K}$ is smaller than $z_K$. 
\\
The oscillator strength of the transverse optical plasmon follows directly from the 
pole-strength of the pole near $\tilde{\omega}_T$ in $\epsilon(\omega)$. From
Eq. \ref{epsilon5} it follows that $S_T$ equals $(\tilde{\omega}_K^2-\tilde{\omega}_T^2)
(\tilde{\omega}_T^2-\tilde{\omega}_I^2)/\tilde{\omega}_T^4$. With the help of this,
it follows that the dependence of $S_T$ on the microscopic parameters $\omega_I$, $\omega_K$, and $K_0n^2$ is 
\begin{equation}
  S_T= \frac
  {(\omega_K^2-\omega_I^2)^2\left(z_Iz_K+2\gamma\right)}
  {(z_I+\frac{2\gamma}{z_K})\omega_K^2+(z_K+\frac{2\gamma}{z_I})\omega_I^2}
 \label{polestrength}
\end{equation}
In Fig. \ref{fig:strength} the plasmon strengths $\tilde{z}_I$, and $\tilde{z}_K$,
and the the pole-strength $S_T$ are displayed as a function of $\gamma$
for a few different sets of $\omega_K$, $\omega_I$, and the volume fraction $z_K$. 
%
%
%
%
\section{Dispersion of the plasma modes}
%
%
%
In the previous section we found for the double layer cuprates {\em two} longitudinal
plasma modes with the electric field vector polarized along the c-direction.
The dispersion of the longitudinal modes
can be calculated, by realizing that for longitudinal modes $\vec{D}=0$ by
virtue of the fact that longitudinally polarized free photons don't exist. Hence in 
Eq. \ref{eom6} the external charge coordinates $Q^e=P^e=0$. The corresponding 
$2\times2$-matrix is easily solved, providing
the two longitudinal branches
\begin{equation}
 \begin{array}{lll}
  \tilde{\omega}_K^2&=&(\frac{1}{2}+\frac{\gamma}{z_K})\omega_{K}^2+
  (\frac{1}{2}+\frac{\gamma}{z_I})\omega_{I}^2+ 
  \sqrt{\left((\frac{1}{2}+\frac{\gamma}{z_K})\omega_{K}^2
     -(\frac{1}{2}+\frac{\gamma}{z_I})\omega_{I}^2\right)^2
   +\frac{(2\gamma\omega_{K}\omega_{I})^2}{z_Kz_I}\cos^2{\frac{k_zd}{2}}}    
  \\
  \tilde{\omega}_I^2&=&(\frac{1}{2}+\frac{\gamma}{z_K})\omega_{K}^2+
  (\frac{1}{2}+\frac{\gamma}{z_I})\omega_{I}^2- 
  \sqrt{\left((\frac{1}{2}+\frac{\gamma}{z_K})\omega_{K}^2
     -(\frac{1}{2}+\frac{\gamma}{z_I})\omega_{I}^2\right)^2
   +\frac{(2\gamma\omega_{K}\omega_{I})^2}{z_Kz_I}\cos^2{\frac{k_zd}{2}}}      
  \\
  \end{array}                                  
 \label{longdisp}
\end{equation}
An example of this dispersion is given in Fig. \ref{lodisp}. In Fig. \ref{sketch} the 
oscillations of these modes are sketched.
%
%
The longitudinal modes in the lefthand side of the figure have a finite
dispersion, provided that $\gamma \neq 0$, in other words, if the electron
gas has finite compressibility. 

In addition to the two longitudinal plasma modes there is {\em one} transverse optical plasma mode, 
and {\em one} transverse polarized pole of the dielectric function at $\omega=0$ 
representing the superconducting dielectric response for fields and currents polarized
along the c-axis, and with the direction of propagation parallel to the planes. In 
the righthand panel of Fig. \ref{sketch} the oscillations of these modes are
sketched.
The transverse mode is coupled to electromagnetic
radiation for long wavelengths, giving rise to coupled plasma-polariton modes
instead of separate photons and transverse optical plasma-modes. The plasma-polariton
dispersion follows from Maxwell's equations in dielectric media, and is given
by the relation between wavenumber $k$ (in the solid) and frequency ($\omega$)
\begin{equation}
  k^2c^2=\epsilon(\omega)\omega^2                                   
\end{equation}
With the dielectric constant given by Eq. \ref{epsilon5} we get
\begin{equation}
  \omega(k)=\left\{
            \left(\frac{\tilde{\omega}_I^2+\tilde{\omega}_K^2+k^2c^2}{2}\right)
            \pm
            \sqrt{\left(\frac{\tilde{\omega}_I^2+\tilde{\omega}_K^2+k^2c^2}{2}\right)^2-
            \tilde{\omega}_I^2\tilde{\omega}_K^2-k^2c^2\tilde{\omega}_T^2}
            \right\}^{1/2}                                  
\end{equation}
This dispersion is sketched in Fig. \ref{todisp}.
%
%
We see, that there are two plasma-polariton branches. The lowest
starts at frequency $\omega_I$ in the long wavelength limit (small k), and quickly merges
with the transverse optical plasma frequency $\omega_T$, as the wavelength is reduced below
$c/\omega_T$, which is of the order of a mm if $\omega_T/2\pi$ is of order 300 GHz. At these
shorter wavelengths the character is almost purely the TO-JPR. The upper branch corresponds
to a conventional transverse Josephson plasmon\cite{tachiki} (without the adjective 'optical')
as it merges with the light-line in the short wavelength limit. The lower branch is novel.
It corresponds to a real polarization wave, and these modes can be used to convert electromagnetic
radiation into microscopic currents, or vice versa. \\
Finally it should be added, that in the short wavelength limit, the presence of finite
compressibility gives rise to  non-trivial dependence of $\epsilon(k,\omega)$ on the
wavevector $k$, which for the transverse modes is parallel to the planes. 
This effect becomes prominent on short wavelength's of the order of lattice
spacings. Fig. 2 was sketched on the scale of $k$ of the order of a few inverse millimeter. On
the scale of a few inverse $\AA$ the k-dependence of $\epsilon(k,\omega)$ will give rise
to finite k-dispersion of $\tilde{\omega}_I(k)$, $\tilde{\omega}_T(k)$ , and $\tilde{\omega}_K(k)$. 
\section{Conclusions}
An expression was derived for
the dielectric function of superconducttors consisting of 
two Josephson coupled superconducting layers per unit cell, taking 
into account the effect of finite compressibility of the electron 
fluid. In this model two longitudinal one transverse optical 
Josephson plasma resonance exist. The latter mode appears as a resonance
in the transverse dielectric function, and it couples directly
to the electric field vector of infrared radiation. A shift
of all plasma frequencies, and a reduction of the intensity of the
transverse optical Josephson plasmon is shown to result from the finite 
compressibility of the electron fluid. 
\section{Acknowledgements}
We like to thank L.N. Bulaevksii for drawing our attention to the dispersion term.
%
%

%
\newpage 
\newpage 
\begin{figure}[h!]
  \centerline{\epsfig{figure=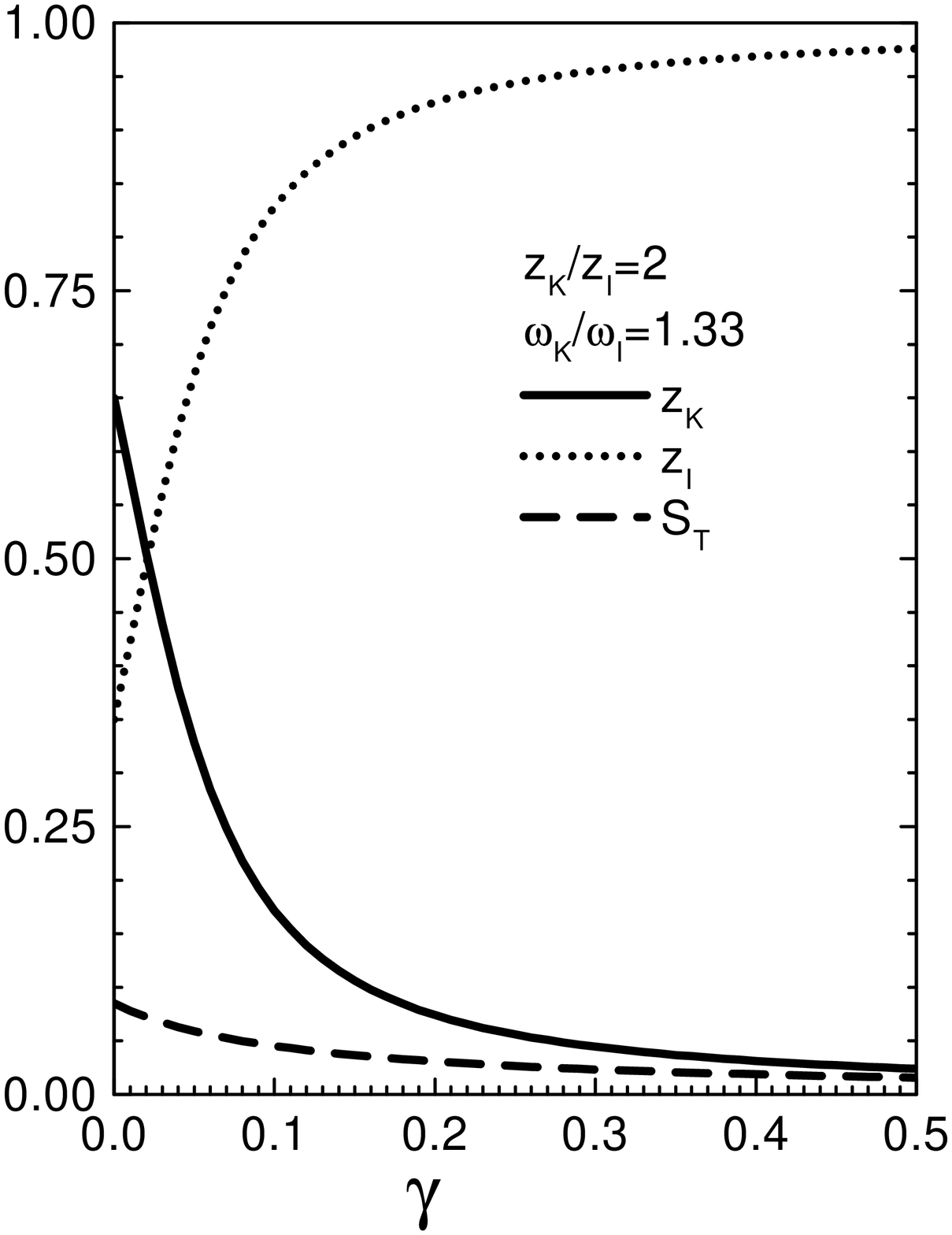,width=7cm,clip=}
  \epsfig{figure=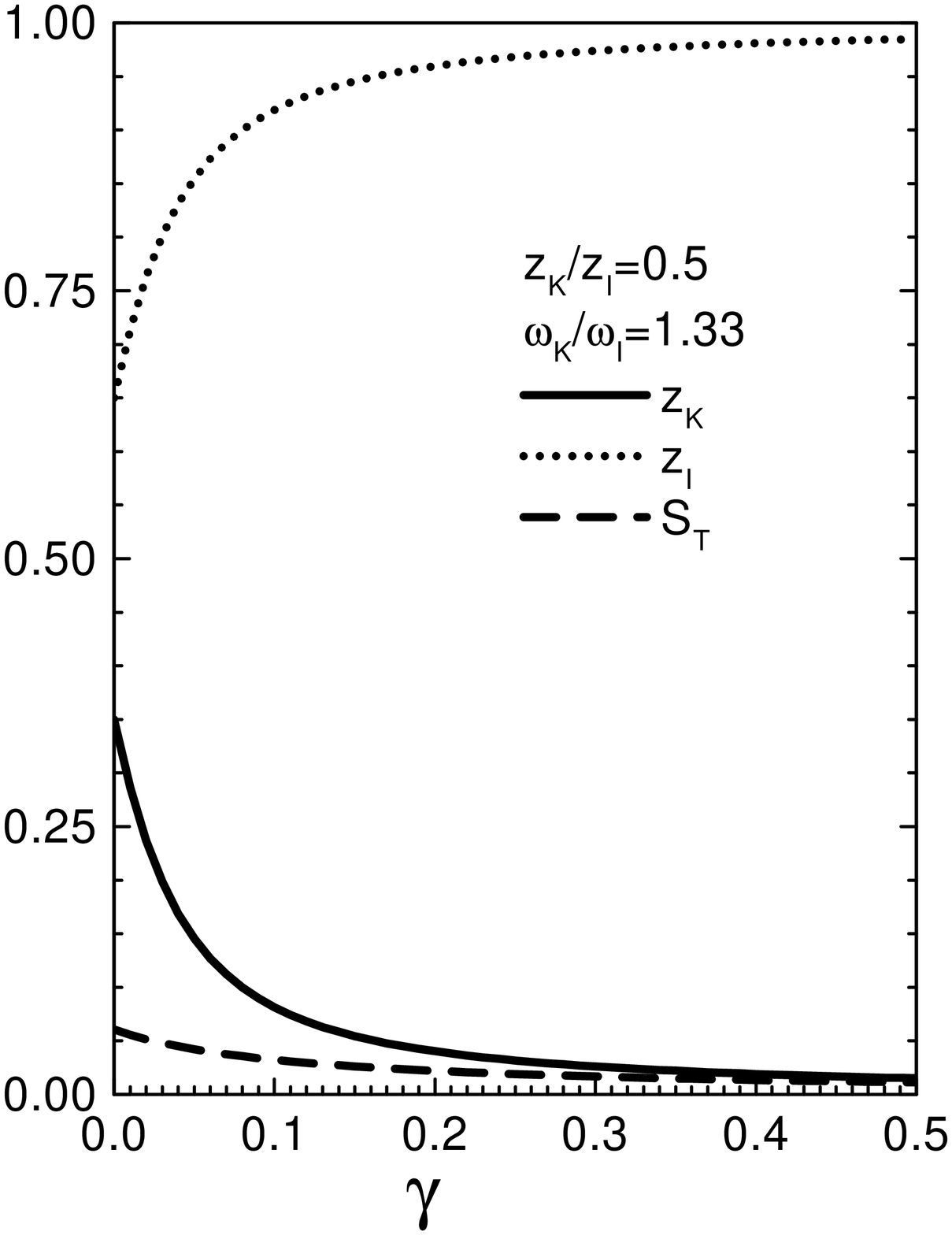,width=7cm,clip=}}
 \caption{Oscillator strength's as a function of $\gamma$. }
 \label{fig:strength'}
\end{figure}
\newpage 
\begin{figure}[h!]
 \centerline{\epsfig{figure=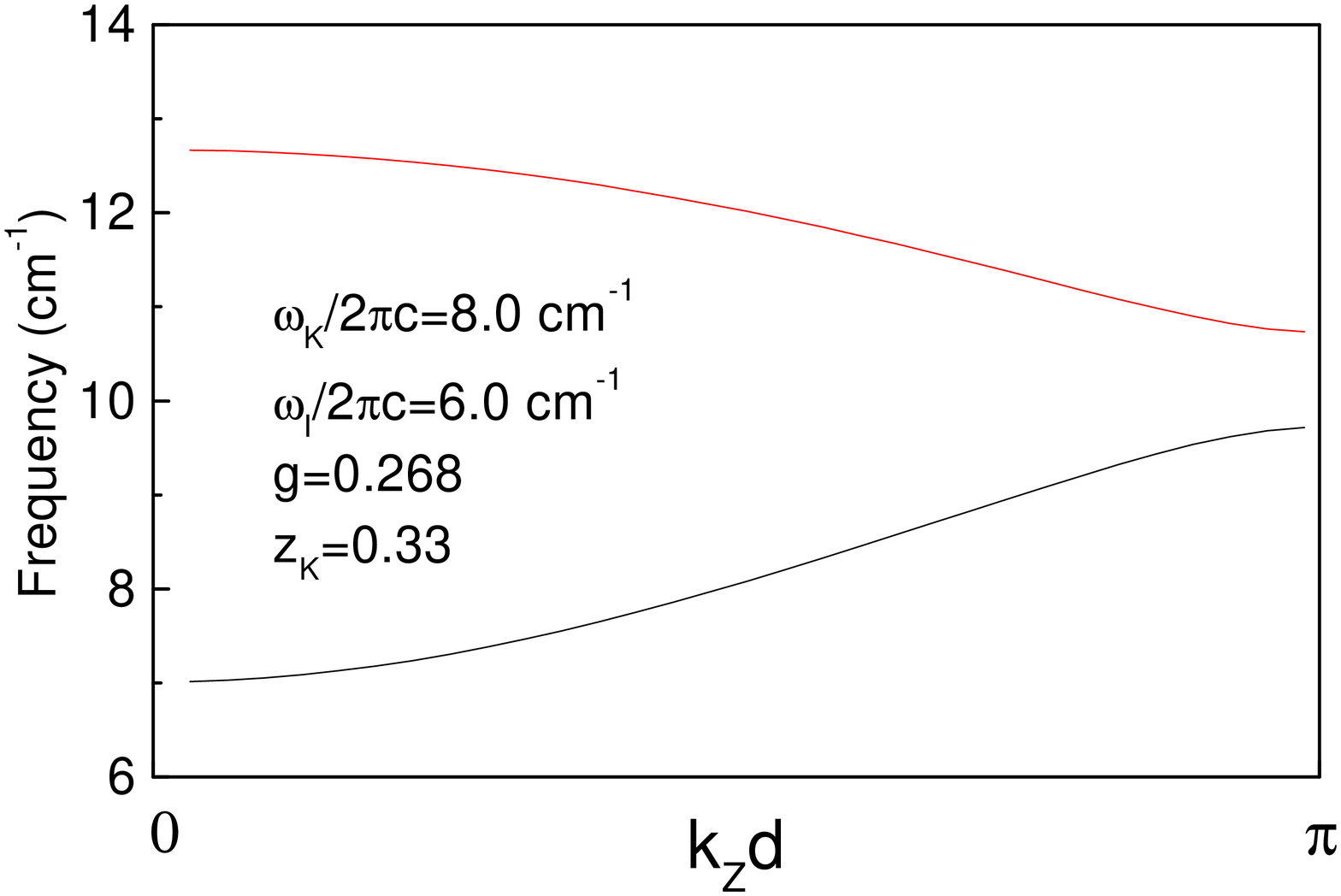,width=14cm,clip=}}
 \caption{Dispersion of the longitudinal JPR's.}
 \label{lodisp}
\end{figure}
\newpage 
\begin{figure}[h!]
 \centerline{\epsfig{figure=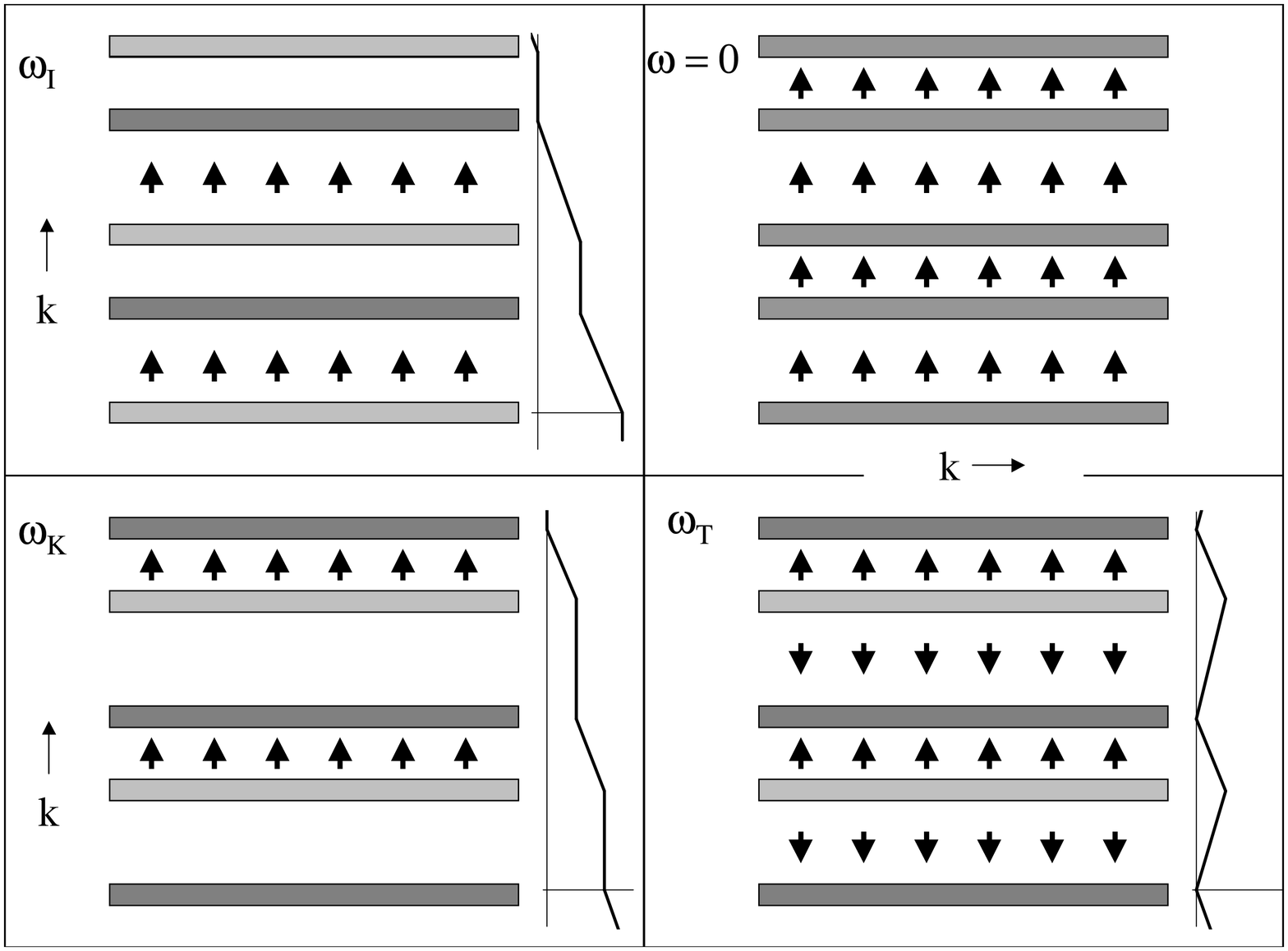,width=10cm,clip=}}
 \caption{Snapshot of the currents (arrows) and planar charge fluctuation
 amplitudes (indicated by gray-scales) of the two sets of transverse and
 longitudinal modes with polarization along the c-direction. On the righthand
 side of each plot the voltage distribution is indicated.}
 \label{sketch}
\end{figure}
\newpage 
\begin{figure}[h!]
 \centerline{\epsfig{figure=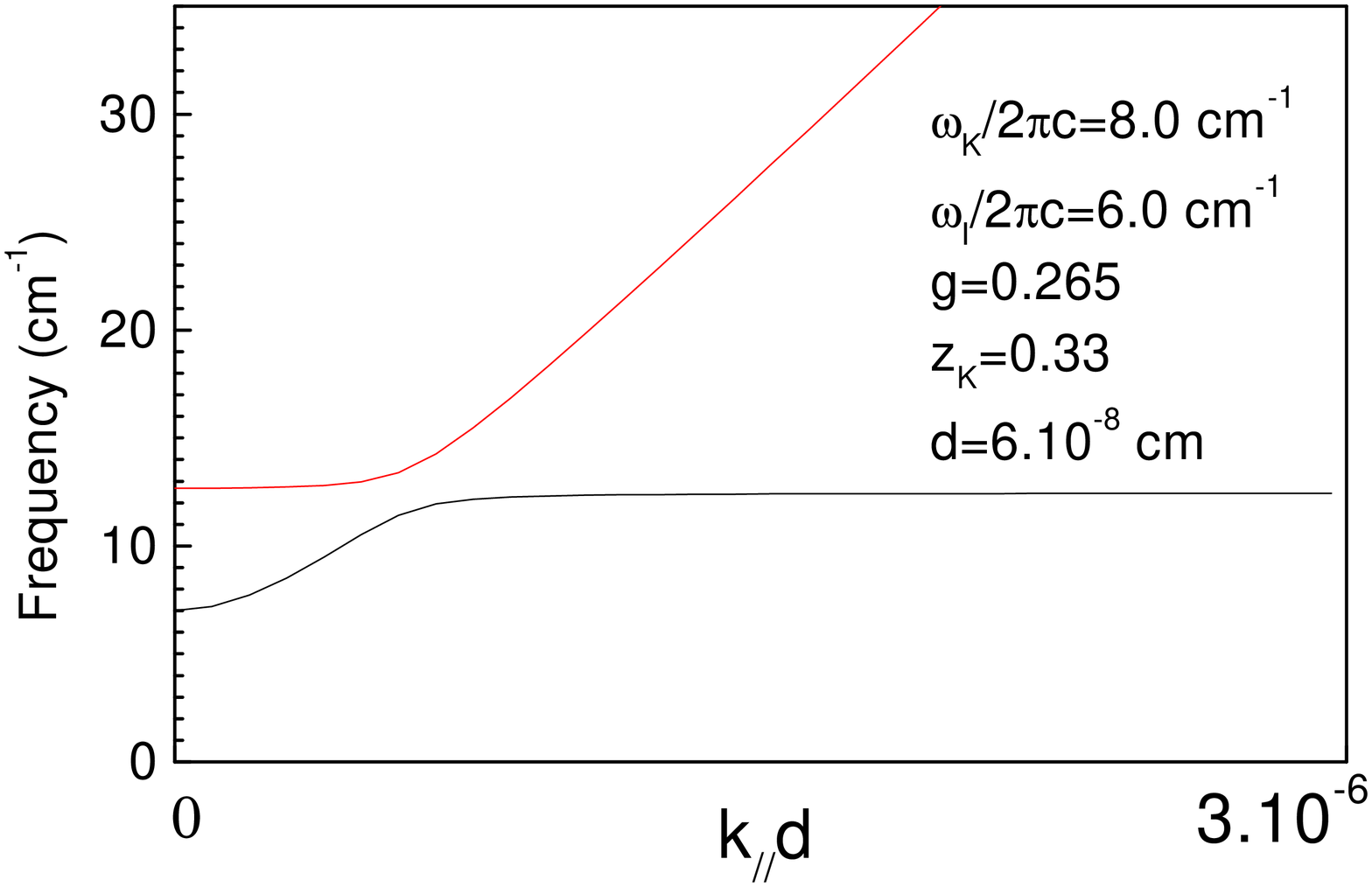,width=14cm,clip=}}
 \caption{Dispersion of the transverse JPR's for small
 values of $k_{\parallel}$ (about one millionth of the Brillouinzone).}
 \label{todisp}
\end{figure}

\begin{references}
\bibitem{howepsilon6} The order of presentation of the main results is not the order in which
the expressions are derived at this point: To derive Eq. \ref{epsilon6}
we used Eqs. \ref{IK} and \ref{epsilon5} to express 
$\tilde{\omega}_T$ in terms of $\gamma$, $\tilde{\omega}_I$, and $\tilde{\omega}_K$. This
is inserted in Eq.\ref{zks1}, leading to Eq. \ref{epsilon6}.
\end{references}
\end{document}